\documentclass[aps,prl,twocolumn]{revtex4}
\usepackage{graphicx}
\usepackage{dcolumn}
\usepackage{amsmath}
\usepackage{amssymb}
\usepackage{subfigure,amsmath,verbatim,moreverb,bm}
\def\be{\begin{equation}}
\def\ee{\end{equation}}
\def\ber{\begin{eqnarray}}
\def\eer{\end{eqnarray}}

\def\rv{{\bf r}}

\def\fv{{\bf f}}

\begin{document}
\title{Strong correlation in Kohn-Sham density functional theory}
\author{Francesc Malet and Paola Gori-Giorgi}
\affiliation{Department of Theoretical Chemistry and Amsterdam Center for Multiscale Modeling, FEW, Vrije Universiteit, De Boelelaan 1083, 1081HV Amsterdam, The Netherlands}
\date{\today}
\begin{abstract}
We 	use the exact strong-interaction limit of the Hohenberg-Kohn energy density functional to approximate the exchange-correlation energy of the restricted Kohn-Sham scheme. Our approximation corresponds to a highly non-local density functional whose functional derivative can be easily constructed, thus transforming {\em exactly}, in a physically transparent way, an important part of the electron-electron interaction into an effective local one-body potential.  We test our approach on quasi-one-dimensional systems, showing that it captures essential features of strong correlation that restricted Kohn-Sham calculations using the currently available approximations cannot describe.
\end{abstract}
\maketitle

In principle, Kohn-Sham (KS) density functional theory (DFT) \cite{HohKoh-PR-64,KohSha-PR-65} should yield the exact ground-state density and energy of {\em any} many-electron system, including physical situations in which electronic correlation is very strong, representing them in terms of {\em non-interacting} electrons. Currently available approximations for KS DFT, however, fail at  properly describing systems approaching the Mott insulating regime \cite{AniZaaAnd-PRB-91}, the breaking of the chemical bond \cite{GruGriBae-JCP-03,CohMorYan-SCI-08}, and localization in low-density nanodevices \cite{GhoGucUmrUllBar-PRB-07,BorTorKosManAbeRei-IJQC-05,AbePolXiaTos-EJPB-07}, to name a few examples (for a recent review, see \cite{CohMorYan-CR-12}). Artificially breaking the spin (or other) symmetry can mimic some (but not all) strong-correlation effects, at the price of a wrong characterization of several properties and of a partial loosening of the rigorous KS DFT framework.

Indeed, it is very counterintuitive that strongly-correlated systems, in which the electron-electron repulsion plays a prominent role, can be exactly represented in terms of non-interacting electrons. For this reason, several authors \cite{Ver-PRL-08,HelTokRub-JCP-09,TemMarMai-JCTC-09,TeaCorHel-JCP-09,TeaCorHel-JCP-10,KurSteKhoVerGro-PRL-10,SteKur-PRL-11,KarPriVer-PRL-11,StoWagWhiBur-PRL-12,BerLiuBurSta-PRL-12,RamGod-PRL-12,BuiBaeSni-PRA-89,FilUmrTau-JCP-94,GrivanBae-PRA-95,GrivanBae-JCP-96,ColSav-JCP-99}   have used accurate many-body solutions of prototypical strongly-correlated systems to obtain (by inversion) and characterize the exact non-interacting KS system. The exact properties needed to describe strong correlation in KS DFT have also been set in a transparent framework \cite{CohMorYan-SCI-08,MorCohYan-PRL-09}. These works made it all the more evident how difficult it is to find adequate approximations of the exact KS system, so that, albeit theoretically possible, it may seem unrealistic to describe strongly-correlated systems with KS DFT \cite{CohMorYan-CR-12}.

Here, we address this skepticism by showing that the strong-interaction limit of the Hohenberg-Kohn (HK) energy density functional yields approximations capturing strong-correlation effects within the non-interacting restricted self-consistent KS scheme. 

The Letter is organized as follows. First, we introduce the formalism, using the strong-interaction limit of the HK functional to transform {\em exactly} an important part of the many-body interaction into an effective local one-body potential, in a physically transparent way.
We then present pilot self-consistent Kohn-Sham calculations, showing that this potential is indeed able to capture strong-correlation effects way beyond the reach of present KS DFT approximations. As a prototypical example, we look at the ``$2k_F\to 4 k_F$'' crossover of electrons confined in quasi-one dimension (Q1D). This crossover is entirely due to the dominant particle-particle repulsion that tends to localize the charge density, destroying the non-interacting shell structure, as it happens in many strong-correlation phenomena. The interest of these results goes beyond quasi-one-dimensional systems, because the latter are a valid test lab for three-dimensional DFT, as clearly discussed in \cite{WagStoBurWhi-PCCP-12}.
Our approximation turns out to be qualitatively right, and quantitatively very accurate for the ionization energies, although less accurate for the ground-state density. We thus conclude by discussing the  inclusion of higher-order corrections and strategies for extending the self-consistent calculations to two and three dimensions. Hartree (effective) atomic units are used throughout. 

{\it Strong-interaction limit--} In Hohenberg and Kohn's formulation \cite{HohKoh-PR-64} the ground-state density  and energy of a many-electron system are obtained by minimizing with respect to the density $\rho(\rv)$ the energy density functional
\be\label{EnergyFunctional}
E[\rho] = F[\rho]+\int d\rv\, v_{\rm ext}(\rv)\,\rho(\rv),
\ee
where $v_{\rm ext}(\rv)$ is the external potential  and $F[\rho]$ is a universal functional of the density, defined as the minimum of the internal energy (kinetic energy $\hat T$ plus electron-electron repulsion $\hat V_{ee}$) with respect to all the fermionic wave functions $\Psi$ that yield the density $\rho(\rv)$ \cite{Lev-PNAS-79},
\be
F[\rho]=\min_{\Psi\to\rho}\langle\Psi|\hat T+\hat V_{ee}|\Psi\rangle.
\label{eq_HK}
\ee
To capture the fermionic nature of the electronic density, Kohn and Sham \cite{KohSha-PR-65} introduced the functional $T_s[\rho]$ by minimizing the expectation value of  $\hat T$  alone over all the fermionic wave functions yielding the given $\rho$ \cite{Lev-PNAS-79},
\be
T_s[\rho]=\min_{\Psi\to\rho}\langle\Psi|\hat T|\Psi\rangle,
\label{eq_Ts}
\ee
thus introducing a reference non-interacting system with the same density of the physical, interacting, one.
The remaining parts of $F[\rho]$, defining the Hartree and exchange-correlation functional, $E_{\rm Hxc}[\rho]\equiv F[\rho]-T_s[\rho]$, are approximated. 

The strong-interaction limit of $F[\rho]$ is given by the functional $V_{ee}^{\rm SCE}[\rho]$, defined as \cite{Sei-PRA-99,SeiPerLev-PRA-99,SeiPerKur-PRL-00,SeiGorSav-PRA-07}
\be
V_{ee}^{\rm SCE}[\rho]=\min_{\Psi\to\rho}\langle\Psi|\hat V_{ee}|\Psi\rangle,
\label{eq_VeeSCE}
\ee
where the acronym ``SCE'' stands for ``strictly-correlated electrons''  \cite{Sei-PRA-99}. The functional $V_{ee}^{\rm SCE}[\rho]$ is the minimum of the electronic interaction alone over all the wave functions yielding the given density. It has been first studied in the seminal work of Seidl and coworkers \cite{Sei-PRA-99,SeiPerLev-PRA-99,SeiPerKur-PRL-00}, and later formalized and evaluated exactly in Refs.~\onlinecite{SeiGorSav-PRA-07,GorSei-PCCP-10,RasSeiGor-PRB-11,ButDepGor-PRA-12}. 

More recently, it has been suggested that a ``SCE DFT'', in which the functional $F[\rho]$ is decomposed as $F[\rho]=V_{ee}^{\rm SCE}[\rho]+E_{kd}[\rho]$  \cite{GorSeiVig-PRL-09,LiuBur-JCP-09,GorSei-PCCP-10}, and the so-called kinetic-decorrelation energy $E_{kd}[\rho]$ is approximated, could be a complementary alternative to KS DFT for systems in which the electron-electron repulsion largely dominates over the electronic kinetic energy. Indeed, SCE DFT works well for low-density many-particle scenarios \cite{GorSeiVig-PRL-09,GorSei-PCCP-10}, but it requires that one knows {\em a priori} that the system is in the strong-interaction regime, and it fails when the fermionic shell structure plays a role \cite{GorSei-PCCP-10}. It also misses several appealing features of KS DFT, e.g. the possibility  to yield (at least in principle) the exact ionization energy. More generally,  orbitals and orbital energies, crucial for chemistry and solid state physics, are totally absent in SCE DFT. 

{\it SCE as a functional for KS DFT --} To combine the advantages of KS DFT and SCE DFT,  here we use the functional $V_{ee}^{\rm SCE}[\rho]$ to approximate  $E_{\rm Hxc}[\rho]$,
\be
E_{\rm Hxc}[\rho]\approx V_{ee}^{\rm SCE}[\rho].
\label{eq_EHxcSCE}
\ee
Equation~\eqref{eq_EHxcSCE} amounts to approximating the constrained minimization over $\Psi$ in the HK functional  \eqref{eq_HK} with the sum of two constrained minima,
\be
\min_{\Psi\to\rho}\langle\Psi|\hat T+\hat V_{ee}|\Psi\rangle\approx \min_{\Psi\to\rho}\langle\Psi|\hat T|\Psi\rangle+\min_{\Psi\to\rho}\langle\Psi|\hat V_{ee}|\Psi\rangle.
\label{eq_FKSSCE}
\ee
This new ``KS SCE'' approach treats both the kinetic energy and the electron-electron repulsion on the same footing. Standard KS DFT emphasizes the non-interacting shell structure, treated accurately  through the functional $T_s[\rho]$, but it misses the features of strong correlation. SCE DFT is biased towards localized ``Wigner-like'' structures in the density, missing the fermionic shell structure.   
Many interesting systems lie in between these two limits, and their complex behavior arises precisely from the competition between the fermionic structure embodied in the kinetic energy and correlation effects due to the electron-electron repulsion. By letting these factors compete in a self-consistent KS procedure, one might be able to get at least a qualitative description of several complex phenomena, amenable to improvement by corrections in the same spirit of standard KS DFT. 

{\it General features of KS SCE --} 
First, notice that for a given density $\rho$, the right-hand side of Eq.~\eqref{eq_FKSSCE} is always less or equal than the left-hand side. Even if minimizing  our energy functional with respect to the density will not yield the exact $\rho$ [as Eq.~\eqref{eq_FKSSCE} is an approximation], it is easy to prove that our final total energy is a lower bound to the exact one.

From the scaling properties \cite{LevPer-PRA-85} of $F[\rho]$, $T_s[\rho]$ and $V_{ee}^{\rm SCE}[\rho]$ it derives that the approximation of Eq.~\eqref{eq_FKSSCE} is accurate both in the weak- and in the strong-interaction limits, while probably  less precise in between.  
By defining, for electrons in $D$ dimensions, $\rho_\gamma(\rv)\equiv\gamma^D\rho(\gamma \rv)$, where $\gamma\ge 0$, we have $T_s[\rho_\gamma]=\gamma^2T_s[\rho]$, $V_{ee}^{\rm SCE}[\rho_\gamma]=\gamma V_{ee}^{\rm SCE}[\rho]$ \cite{GorSei-PCCP-10}, and $F[\rho_\gamma]=\gamma^2F_{1/\gamma}[\rho]$, where $F_{1/\gamma}[\rho]$ means that the Coulomb coupling constant in $F[\rho]$ is rescaled with $1/\gamma$. It is then easy to see that both sides of Eq.~\eqref{eq_FKSSCE} tend to $T_s[\rho_\gamma]$ when $\gamma\to\infty$ (high density or weak interaction) and to $V_{ee}^{\rm SCE}[\rho_\gamma]$ when $\gamma\to 0$ (low density or strong interaction). 

Since KS SCE tends to the exact density and energy in the strong-interaction limit, the corresponding KS potential should have the features that are expected for a KS description of strong correlation \cite{HelTokRub-JCP-09,BuiBaeSni-PRA-89,StoWagWhiBur-PRL-12}. We discuss first why, physically and mathematically, the SCE potential [Eqs.~\eqref{eq_vSCE}-\eqref{eq_funcder} below] is expected to have these features, which we then test practically with self-consistent calculations in Q1D. 

Physically, the functional $V_{ee}^{\rm SCE}[\rho]$ portrays the  {\em strict correlation} regime, where the position $\rv$ of one electron determines all the other $N-1$ electronic positions $\rv_i$ through the so-called {\em co-motion functions}, $\rv_i=\fv_i[\rho](\rv)$, some non-local functionals of the density \cite{SeiGorSav-PRA-07,GorVigSei-JCTC-09,GorSeiVig-PRL-09,ButDepGor-PRA-12}. Therefore, the net repulsion on an electron at position ${\bf r}$ due to the other $N-1$ electrons depends on $\rv$ alone. Its effect can then be {\it exactly} represented \cite{SeiGorSav-PRA-07,GorSeiVig-PRL-09,ButDepGor-PRA-12} by a local one-body potential \footnote{We have defined $\tilde{v}_{\rm SCE}[\rho](\rv)=-v_{\rm SCE}[\rho](\rv)$ of Refs.~\onlinecite{SeiGorSav-PRA-07,GorVigSei-JCTC-09,GorSeiVig-PRL-09,ButDepGor-PRA-12}, as here we seek an effective potential for KS theory, corresponding to the net electron-electron repulsion acting on an electron at position $\rv$, while in Refs.~\onlinecite{SeiGorSav-PRA-07,GorVigSei-JCTC-09,GorSeiVig-PRL-09,ButDepGor-PRA-12}  an effective potential for the SCE system was sought, {\it i.e.}, the potential that {\it compensates} the net electron-electron repulsion.},
\be
\nabla \tilde{v}_{\rm SCE}[\rho](\rv)=-\sum_{i= 2}^N \frac{\rv-\fv_i[\rho](\rv)}{|\rv-\fv_i[\rho](\rv)|^3}.
\label{eq_vSCE}
\ee
The physical meaning of Eq.~\eqref{eq_vSCE} is very transparent: at each position $\rv$, $\nabla\tilde{v}_{\rm SCE}[\rho](\rv)$ exerts the same force as the net electron-electron repulsion. We also have \cite{GorSeiVig-PRL-09,ButDepGor-PRA-12}
\be
\frac{\delta V_{ee}^{\rm SCE}[\rho]}{\delta \rho(\rv)}=\tilde{v}_{\rm SCE}[\rho](\rv),
\label{eq_funcder}
\ee
so that the approximation of Eq.~\eqref{eq_EHxcSCE} corresponds to model the exchange-correlation potential $v_{xc}[\rho](\rv)$ of KS DFT as $v_{xc}[\rho](\rv)\approx \tilde{v}_{\rm SCE}[\rho](\rv)-v_H[\rho](\rv)$, where $v_H[\rho](\rv)$ is the Hartree potential. The functional $V_{ee}^{\rm SCE}[\rho]$, being essentially a classical repulsion energy, favors localized charge densities.  
When evaluated with a {\em delocalized} density $\rho$, its functional derivative \eqref{eq_funcder} as a function of $\rv$ displays strong variations pushing electrons towards localization.
Otherwise said, Eqs.~\eqref{eq_vSCE}-\eqref{eq_funcder} transfer the effects of strong-correlation into a physically meaningful, effective local potential, expressed as the functional derivative of a rigorous KS density functional.

While KS SCE does not use explicitly the Hartree functional, the correct electrostatics is still captured, since $V_{ee}^{\rm SCE}[\rho]$ is the classical electrostatic minimum in the given density $\rho$. 
Moreover, the potential $\tilde{v}_{\rm SCE}[\rho](\rv)$  stems from a wave function (the SCE one \cite{SeiGorSav-PRA-07,GorVigSei-JCTC-09}) and is therefore completely self-interaction free. Similarly, we expect $\tilde{v}_{\rm SCE}[\rho](\rv)$ to have a derivative discontinuity that will be analyzed elsewhere \cite{MirSeiGor-XXX}.
\begin{figure}
\includegraphics[width=7cm]{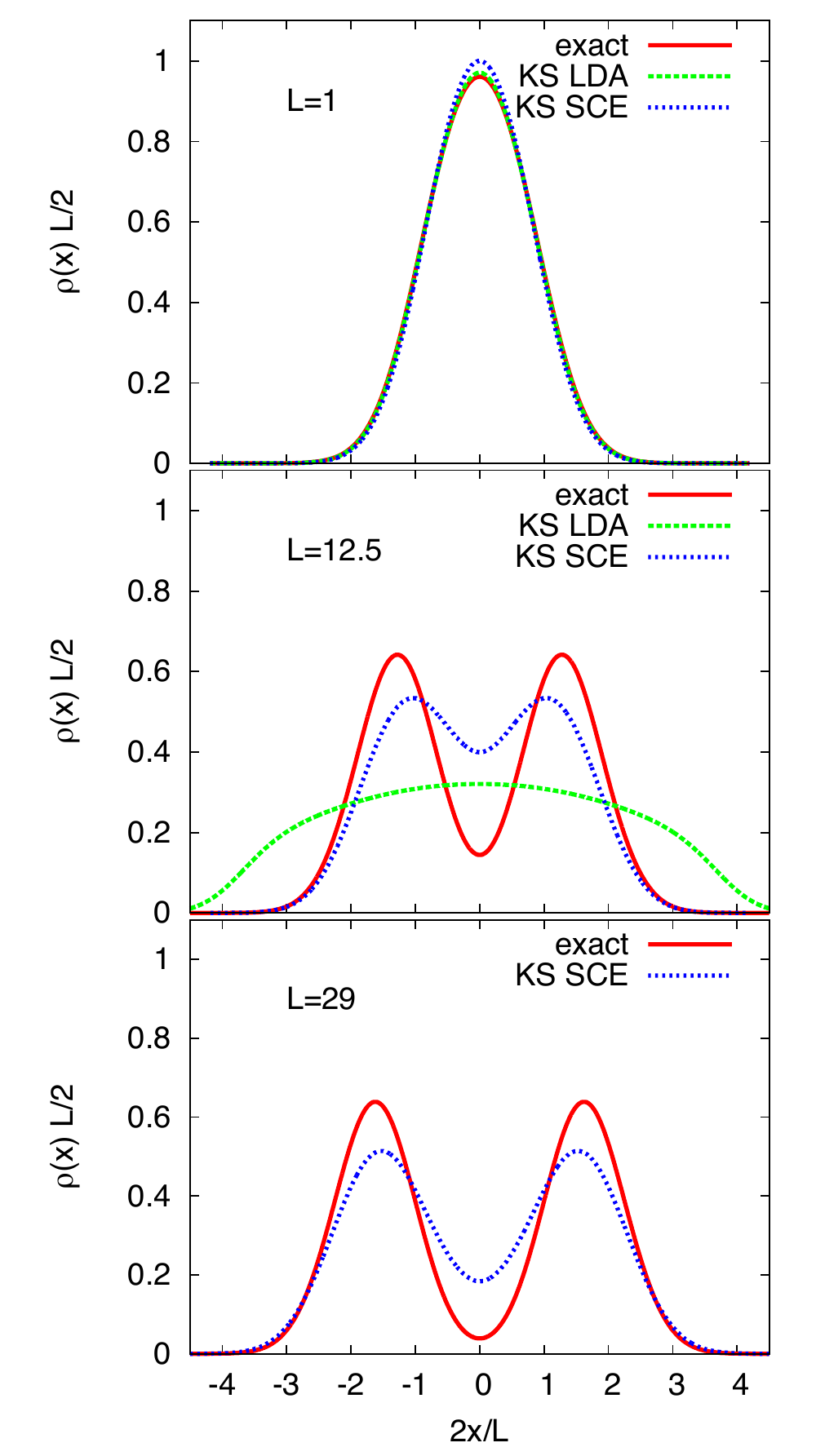}
   \caption{(color online) Self-consistent densities for $N=2$ electrons in Q1D [hamiltonian of Eqs.~\eqref{eq_H1D}-\eqref{eq_int1D} with $b=0.1$ and $v_{\rm ext}(x)=\frac{1}{2}\omega^2x^2$], in units of the effective confinement length $L=2 \omega^{-1/2}$ (here and in the following figures). The exact results are compared with  KS LDA and KS SCE approximations. At large $L$ the KS LDA calculations do not converge, while KS SCE approaches the exact solution.}
\label{fig_densN2}
\end{figure}

{\it Self-consistent KS SCE calculations in Q1D --} As a pilot test of the approximation of Eq.~\eqref{eq_FKSSCE}, we consider $N$ electrons in a thin quantum wire with hamiltonian
\be
\hat{H}=-\frac{1}{2}\sum_{i=1}^N \frac{\partial^2}{\partial x_i^2}+\sum_{i=1}^{N-1}\sum_{j=i+1}^N w_b(|x_i-x_j|)+\sum_{i=1}^N v_{\rm ext}(x_i),
\label{eq_H1D}
\ee 
where the effective Q1D interaction is obtained by integrating the Coulomb repulsion on the lateral degrees of freedom \cite{BedSzaChwAda-PRB-03},
\be
w_b(x)=\frac{\sqrt{\pi}}{2\,b}\,\exp\left(\frac{x^2}{4\,b^2}\right){\rm erfc}\left(\frac{x}{2\,b}\right).
\label{eq_int1D}
\ee
The parameter $b$ fixes the thickness of the wire, and ${\rm erfc}(x)$ is the complementary error function.
The interaction $w_b(x)$ has a coulombic tail, $w_b(x\to\infty)=1/x$, and is finite at the origin, where it has a cusp.

The co-motion functions $f_i(x)$ for $N$ electrons can be constructed from the density $\rho(x)$ \cite{Sei-PRA-99,RasSeiGor-PRB-11,ButDepGor-PRA-12}: 
\be
f_i(x)=
\Bigg\{
\begin{array}{l}
N_e^{-1}[N_e(x)+i-1] \qquad \qquad \; x\leq a_{N+1-i} \\
 N_e^{-1}[N_e(x)+i-1-N] \qquad x> a_{N+1-i},
\end{array} 
\ee
where the function $N_e(x)$ is
\be
N_e(x)=\int_{-\infty}^x\rho(x')\,dx',
\ee
and $a_{k}=N_e^{-1}(k)$. Equation \eqref{eq_vSCE} becomes in this case
\be
\tilde{v}'_{\rm SCE}[\rho](x)=\sum_{i=2}^N w_b'(|x-f_i(x)|){\rm sgn}(x-f_i(x)).
\label{eq_vSCE1D}
\ee
We  then solve self-consistently the restricted KS equations in the KS potential $v_{\rm KS}(x)=v_{\rm ext}(x)+\tilde{v}_{\rm SCE}[\rho](x)$, where $\tilde{v}_{\rm SCE}[\rho](x)$ is obtained by integrating Eq.~\eqref{eq_vSCE1D} with the boundary condition $\tilde{v}_{\rm SCE}[\rho](|x|\to\infty)=0$.
\begin{figure}
   \includegraphics[width=7.5cm]{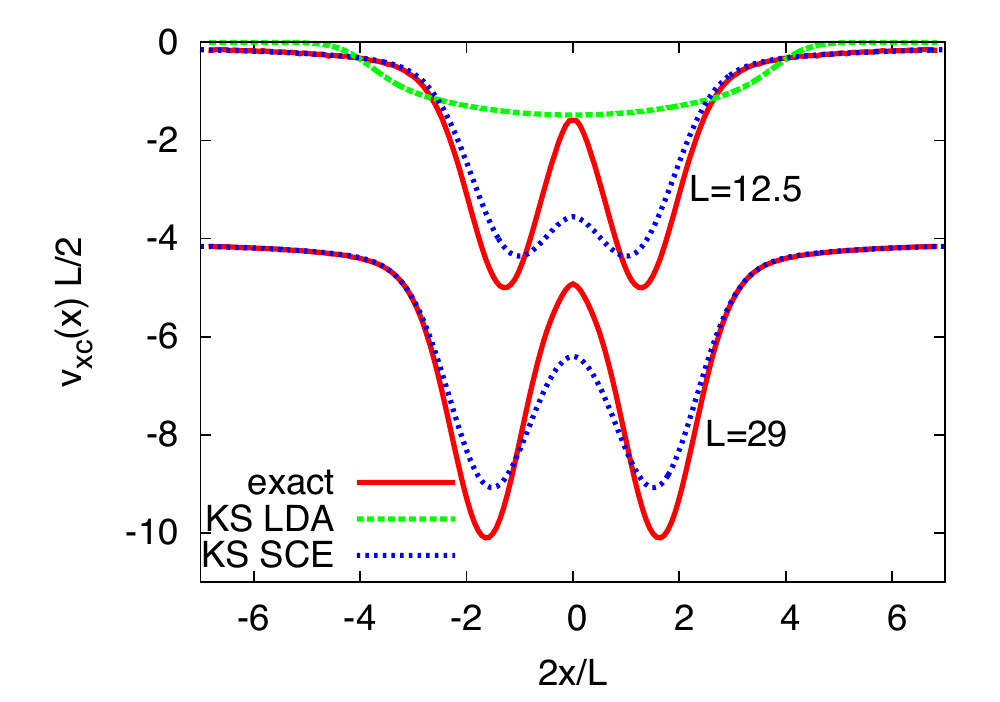}
   \caption{(color online) Self-consistent exchange-correlation potentials for the same system of Fig.~\ref{fig_densN2}. For clarity, the potentials for $L=29$ have been shifted by $-4$.}
\label{fig_vxc}
\end{figure}

\begin{figure}
   \includegraphics[width=7cm]{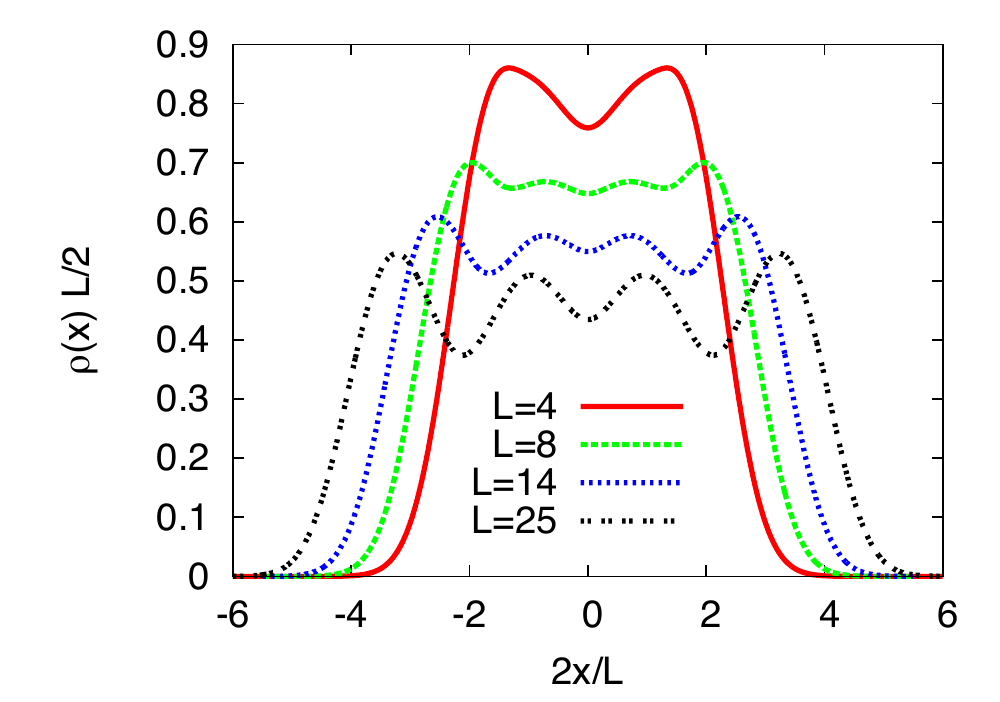}
   \caption{(color online) Self-consistent KS SCE densities for $N=4$ electrons in Q1D [hamiltonian of Eqs.~\eqref{eq_H1D}-\eqref{eq_int1D} with $b=0.1$ and $v_{\rm ext}(x)=\frac{1}{2}\omega^2x^2$].}
\label{fig_densN4}
\end{figure}

Here, we aim at showing that this KS SCE approach captures essential features of strong correlation out of reach for standard restricted KS calculations. A simple but very representative example is provided by  Abedinpour {\it et. al.} \cite{AbePolXiaTos-EJPB-07}, who considered the external harmonic confinement $v_{\rm ext}(x)=\frac{1}{2}\omega^2x^2$, and performed self-consistent KS calculations within the local density approximation (LDA, \cite{CasSorSen-PRB-06}).  Fig.~\ref{fig_densN2} shows our results for $N=2$, together with accurate exact values \cite{AbePolXiaTos-EJPB-07}: as expected, KS LDA works well when correlation is weak or moderate, a case characterized by relatively small values of the effective confinement length $L=2\omega^{-1/2}$.
As correlation becomes stronger (large $L$), KS LDA cannot describe the ``$2k_F\to 4 k_F$'' crossover,  simply reflected by the doubling of the number of peaks in the density. Indeed, a local or semilocal functional of the density cannot describe this crossover \cite{AbePolXiaTos-EJPB-07}, and exact exchange performs even worse. To localize the charge density, the self-consistent KS potential must build a ``bump'' (or barrier) between the electrons \cite{AbePolXiaTos-EJPB-07}. This ``bump'' was discussed in Refs.~\onlinecite{BuiBaeSni-PRA-89} and~\onlinecite{HelTokRub-JCP-09}: it is expected to be the key feature enabling a KS DFT description of the Mott transition and the breaking of the chemical bond, and it must be a very non-local effect \cite{HelTokRub-JCP-09}.
We see in Fig.~\ref{fig_densN2} that the self-consistent KS SCE densities, although, as expected, less accurate in between the weak- and the strong-interaction cases, capture the transition to the strongly-correlated regime, thus building, at least partially, the ``bump'' in the self-consistent KS potential. This is confirmed by the exchange-correlation potentials reported in Fig.~\ref{fig_vxc}: we see that the ``bump'' is clearly present and gets closer to the exact one as the strong-interaction regime is approached. The long-range part of the SCE potential is also remarkably accurate, as expected from the fact that the SCE functional is self-interaction free.
 Fig.~\ref{fig_densN4} displays the KS SCE densities for $N=4$ electrons: again, we clearly see the crossover from two peaks (the non-interacting shell structure) to four peaks (charge localization).
\begin{figure}
   \includegraphics[width=7cm]{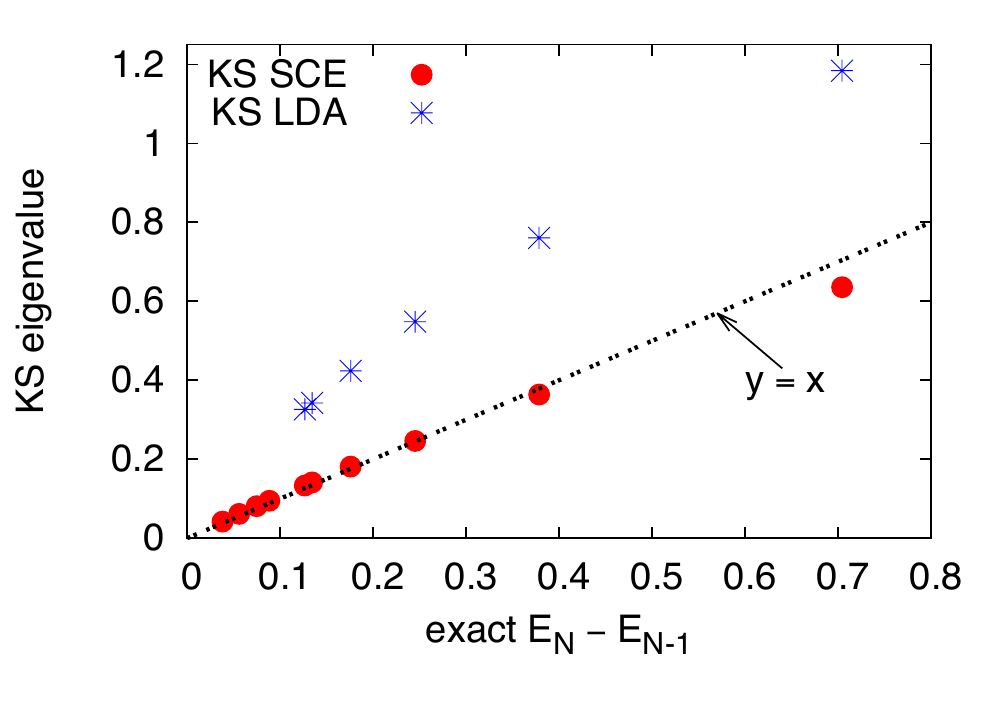}
   \caption{(color online) The KS eigenvalue obtained in the self-consistent KS LDA and KS SCE calculations for the same $N=2$ electron problem considered in Fig.~\ref{fig_densN2}, plotted against the negative of the exact ionization energy, $E_N-E_{N-1}$.  }
\label{fig_ioniz}
\end{figure}

In the exact KS theory, the highest occupied KS eigenvalue is equal to minus the exact ionization potential $I_0=E_{N-1}-E_N$ \cite{AlmBar-PRB-85,LevPerSah-PRA-84}. In Fig.~\ref{fig_ioniz} we plot the KS LDA and KS SCE eigenvalues for $N=2$, as a function of the exact difference $E_N-E_{N-1}$ for several harmonic confinement strengths. We see that KS SCE is remarkably accurate \footnote{It is straightforward to generalize the arguments given in Refs.~\onlinecite{AlmBar-PRB-85} and \onlinecite{LevPerSah-PRA-84} to the harmonic external potential.}.

{\it Conclusions and Perspectives --} The exact strong-interaction limit has the promise of extending KS DFT applicability to strongly-correlated systems, while retaining the appealing properties of the Kohn-Sham approach. In Q1D the computational cost of KS SCE compares to KS LDA. 
Crucial for future applications is calculating $V_{ee}^{\rm SCE}[\rho]$ and $\tilde{v}_{\rm SCE}[\rho](\rv)$ also for general two- and three-dimensional systems. An enticing route towards this goal involves the mass-transportation-theory reformulation of the SCE functional \cite{ButDepGor-PRA-12}, in which $V_{ee}^{\rm SCE}[\rho]$ is given by the maximum of the Kantorovich dual problem,
\be
\max_{u}\left\{ \int u(\rv)\rho(\rv) d\rv \ : \ \sum_{i=1}^N u(\rv_i)\le \sum_{i=1}^{N-1}\sum_{j>i}^N \frac{1}{|\rv_i-\rv_j|} \right\}, \nonumber
\label{eq_VeeKantorovic}
\ee
where $u(\rv)=\tilde{v}_{\rm SCE}[\rho](\rv)+C$, and $C$ is a constant \cite{ButDepGor-PRA-12}. This is a maximization under linear constraints that yields in one shot the functional and its functional derivative, and can also inspire  approximate and simplified approaches to the construction of $V_{ee}^{\rm SCE}[\rho]$ and $\tilde{v}_{\rm SCE}[\rho](\rv)$ \cite{MenLin-arxiv-12}, a critical step for the computational cost of KS SCE.
Another important issue is to add corrections to Eq.~\eqref{eq_EHxcSCE}. One can, more generally, decompose $F[\rho]$ as
\be
F[\rho]=T_s[\rho]+V_{ee}^{\rm SCE}[\rho]+T_c[\rho]+V_{ee}^d[\rho],
\ee
where $T_c[\rho]$ (kinetic correlation energy) is the difference between the true kinetic energy and $T_s[\rho]$, and $V_{ee}^d[\rho]$ (electron-electron decorrelation energy) is the difference between the true expectation of $\hat{V}_{ee}$ and $V_{ee}^{\rm SCE}[\rho]$. A ``first-order'' approximation for $T_c[\rho]+V_{ee}^d[\rho]$ can be, in principle, included exactly using the formalism developed in Ref.~\onlinecite{GorVigSei-JCTC-09}, but other approximations, {\it e.g.} in the spirit of Ref.~\onlinecite{SeiPerKur-PRA-00}, can also be constructed.

{\it Acknowledgments --} We thank J.~Lorenzana and V.~Brosco for inspiring discussions, and K.~J.~H.~Giesbertz, A. Mirtschink, M. Seidl, and G. Vignale for  critical readings of the manuscript. This work was supported by the Netherlands Organization for Scientific Research (NWO) through a Vidi grant.
 

\begin{thebibliography}{46}
\expandafter\ifx\csname natexlab\endcsname\relax\def\natexlab#1{#1}\fi
\expandafter\ifx\csname bibnamefont\endcsname\relax
  \def\bibnamefont#1{#1}\fi
\expandafter\ifx\csname bibfnamefont\endcsname\relax
  \def\bibfnamefont#1{#1}\fi
\expandafter\ifx\csname citenamefont\endcsname\relax
  \def\citenamefont#1{#1}\fi
\expandafter\ifx\csname url\endcsname\relax
  \def\url#1{\texttt{#1}}\fi
\expandafter\ifx\csname urlprefix\endcsname\relax\def\urlprefix{URL }\fi
\providecommand{\bibinfo}[2]{#2}
\providecommand{\eprint}[2][]{\url{#2}}

\bibitem[{\citenamefont{Hohenberg and Kohn}(1964)}]{HohKoh-PR-64}
\bibinfo{author}{\bibfnamefont{P.}~\bibnamefont{Hohenberg}} \bibnamefont{and}
  \bibinfo{author}{\bibfnamefont{W.}~\bibnamefont{Kohn}},
  \bibinfo{journal}{Phys. Rev.} \textbf{\bibinfo{volume}{{136}}},
  \bibinfo{pages}{B 864} (\bibinfo{year}{1964}).

\bibitem[{\citenamefont{Kohn and Sham}(1965)}]{KohSha-PR-65}
\bibinfo{author}{\bibfnamefont{W.}~\bibnamefont{Kohn}} \bibnamefont{and}
  \bibinfo{author}{\bibfnamefont{L.~J.} \bibnamefont{Sham}},
  \bibinfo{journal}{Phys. Rev. A} \textbf{\bibinfo{volume}{140}},
  \bibinfo{pages}{1133} (\bibinfo{year}{1965}).

\bibitem[{\citenamefont{Anisimov et~al.}(1991)\citenamefont{Anisimov, Zaanen,
  and Andersen}}]{AniZaaAnd-PRB-91}
\bibinfo{author}{\bibfnamefont{V.~I.} \bibnamefont{Anisimov}},
  \bibinfo{author}{\bibfnamefont{J.}~\bibnamefont{Zaanen}}, \bibnamefont{and}
  \bibinfo{author}{\bibfnamefont{O.~K.} \bibnamefont{Andersen}},
  \bibinfo{journal}{Phys. Rev. B} \textbf{\bibinfo{volume}{44}},
  \bibinfo{pages}{943} (\bibinfo{year}{1991}).

\bibitem[{\citenamefont{Gr\"uning et~al.}(2003)\citenamefont{Gr\"uning,
  Gritsenko, and Baerends}}]{GruGriBae-JCP-03}
\bibinfo{author}{\bibfnamefont{M.}~\bibnamefont{Gr\"uning}},
  \bibinfo{author}{\bibfnamefont{O.~V.} \bibnamefont{Gritsenko}},
  \bibnamefont{and} \bibinfo{author}{\bibfnamefont{E.~J.}
  \bibnamefont{Baerends}}, \bibinfo{journal}{J. Chem. Phys.}
  \textbf{\bibinfo{volume}{{118}}}, \bibinfo{pages}{7183}
  (\bibinfo{year}{2003}).

\bibitem[{\citenamefont{Cohen et~al.}(2008)\citenamefont{Cohen, Mori-Sanchez,
  and Yang}}]{CohMorYan-SCI-08}
\bibinfo{author}{\bibfnamefont{A.~J.} \bibnamefont{Cohen}},
  \bibinfo{author}{\bibfnamefont{P.}~\bibnamefont{Mori-Sanchez}},
  \bibnamefont{and} \bibinfo{author}{\bibfnamefont{W.~T.} \bibnamefont{Yang}},
  \bibinfo{journal}{Science} \textbf{\bibinfo{volume}{{321}}},
  \bibinfo{pages}{792} (\bibinfo{year}{2008}).

\bibitem[{\citenamefont{Ghosal et~al.}(2007)\citenamefont{Ghosal, Guclu,
  Umrigar, Ullmo, and Baranger}}]{GhoGucUmrUllBar-PRB-07}
\bibinfo{author}{\bibfnamefont{A.}~\bibnamefont{Ghosal}},
  \bibinfo{author}{\bibfnamefont{A.~D.} \bibnamefont{Guclu}},
  \bibinfo{author}{\bibfnamefont{C.~J.} \bibnamefont{Umrigar}},
  \bibinfo{author}{\bibfnamefont{D.}~\bibnamefont{Ullmo}}, \bibnamefont{and}
  \bibinfo{author}{\bibfnamefont{H.~U.} \bibnamefont{Baranger}},
  \bibinfo{journal}{Phys. Rev. B} \textbf{\bibinfo{volume}{76}},
  \bibinfo{pages}{085341} (\bibinfo{year}{2007}).

\bibitem[{\citenamefont{Borgh et~al.}(2005)\citenamefont{Borgh, Toreblad,
  Koskinen, Manninen, Aberg, and Reimann}}]{BorTorKosManAbeRei-IJQC-05}
\bibinfo{author}{\bibfnamefont{M.}~\bibnamefont{Borgh}},
  \bibinfo{author}{\bibfnamefont{M.}~\bibnamefont{Toreblad}},
  \bibinfo{author}{\bibfnamefont{M.}~\bibnamefont{Koskinen}},
  \bibinfo{author}{\bibfnamefont{M.}~\bibnamefont{Manninen}},
  \bibinfo{author}{\bibfnamefont{S.}~\bibnamefont{Aberg}}, \bibnamefont{and}
  \bibinfo{author}{\bibfnamefont{S.~M.} \bibnamefont{Reimann}},
  \bibinfo{journal}{Int. J. Quantum Chem.} \textbf{\bibinfo{volume}{{105}}},
  \bibinfo{pages}{817} (\bibinfo{year}{2005}).

\bibitem[{\citenamefont{Abedinpour et~al.}(2007)\citenamefont{Abedinpour,
  Polini, Xianlong, and Tosi}}]{AbePolXiaTos-EJPB-07}
\bibinfo{author}{\bibfnamefont{S.~H.} \bibnamefont{Abedinpour}},
  \bibinfo{author}{\bibfnamefont{M.}~\bibnamefont{Polini}},
  \bibinfo{author}{\bibfnamefont{G.}~\bibnamefont{Xianlong}}, \bibnamefont{and}
  \bibinfo{author}{\bibfnamefont{M.~P.} \bibnamefont{Tosi}},
  \bibinfo{journal}{Eur. Phys. J. B} \textbf{\bibinfo{volume}{{56}}},
  \bibinfo{pages}{127} (\bibinfo{year}{2007}).

\bibitem[{\citenamefont{Cohen et~al.}(2012)\citenamefont{Cohen, Mori-S\'anchez,
  and Yang}}]{CohMorYan-CR-12}
\bibinfo{author}{\bibfnamefont{A.~J.} \bibnamefont{Cohen}},
  \bibinfo{author}{\bibfnamefont{P.}~\bibnamefont{Mori-S\'anchez}},
  \bibnamefont{and} \bibinfo{author}{\bibfnamefont{W.}~\bibnamefont{Yang}},
  \bibinfo{journal}{Chem. Rev.} \textbf{\bibinfo{volume}{112}},
  \bibinfo{pages}{289} (\bibinfo{year}{2012}).

\bibitem[{\citenamefont{Verdozzi}(2008)}]{Ver-PRL-08}
\bibinfo{author}{\bibfnamefont{C.}~\bibnamefont{Verdozzi}},
  \bibinfo{journal}{Phys. Rev. Lett.} \textbf{\bibinfo{volume}{101}},
  \bibinfo{pages}{166401} (\bibinfo{year}{2008}).

\bibitem[{\citenamefont{Helbig et~al.}(2009)\citenamefont{Helbig, Tokatly, and
  Rubio}}]{HelTokRub-JCP-09}
\bibinfo{author}{\bibfnamefont{N.}~\bibnamefont{Helbig}},
  \bibinfo{author}{\bibfnamefont{I.~V.} \bibnamefont{Tokatly}},
  \bibnamefont{and} \bibinfo{author}{\bibfnamefont{A.}~\bibnamefont{Rubio}},
  \bibinfo{journal}{J. Chem. Phys.} \textbf{\bibinfo{volume}{131}},
  \bibinfo{pages}{224105} (\bibinfo{year}{2009}).

\bibitem[{\citenamefont{Tempel et~al.}(2009)\citenamefont{Tempel, Mart\'inez,
  and Maitra}}]{TemMarMai-JCTC-09}
\bibinfo{author}{\bibfnamefont{D.~G.} \bibnamefont{Tempel}},
  \bibinfo{author}{\bibfnamefont{T.~J.} \bibnamefont{Mart\'inez}},
  \bibnamefont{and} \bibinfo{author}{\bibfnamefont{N.~T.}
  \bibnamefont{Maitra}}, \bibinfo{journal}{J. Chem. Theory Comput.}
  \textbf{\bibinfo{volume}{5}}, \bibinfo{pages}{770} (\bibinfo{year}{2009}).

\bibitem[{\citenamefont{Teale et~al.}(2009)\citenamefont{Teale, Coriani, and
  Helgaker}}]{TeaCorHel-JCP-09}
\bibinfo{author}{\bibfnamefont{A.~M.} \bibnamefont{Teale}},
  \bibinfo{author}{\bibfnamefont{S.}~\bibnamefont{Coriani}}, \bibnamefont{and}
  \bibinfo{author}{\bibfnamefont{T.}~\bibnamefont{Helgaker}},
  \bibinfo{journal}{J. Chem. Phys.} \textbf{\bibinfo{volume}{{130}}},
  \bibinfo{pages}{104111} (\bibinfo{year}{2009}).

\bibitem[{\citenamefont{Teale et~al.}(2010)\citenamefont{Teale, Coriani, and
  Helgaker}}]{TeaCorHel-JCP-10}
\bibinfo{author}{\bibfnamefont{A.~M.} \bibnamefont{Teale}},
  \bibinfo{author}{\bibfnamefont{S.}~\bibnamefont{Coriani}}, \bibnamefont{and}
  \bibinfo{author}{\bibfnamefont{T.}~\bibnamefont{Helgaker}},
  \bibinfo{journal}{J. Chem. Phys.} \textbf{\bibinfo{volume}{{132}}},
  \bibinfo{pages}{164115} (\bibinfo{year}{2010}).

\bibitem[{\citenamefont{Kurth et~al.}(2010)\citenamefont{Kurth, Stefanucci,
  Khosravi, Verdozzi, and Gross}}]{KurSteKhoVerGro-PRL-10}
\bibinfo{author}{\bibfnamefont{S.}~\bibnamefont{Kurth}},
  \bibinfo{author}{\bibfnamefont{G.}~\bibnamefont{Stefanucci}},
  \bibinfo{author}{\bibfnamefont{E.}~\bibnamefont{Khosravi}},
  \bibinfo{author}{\bibfnamefont{C.}~\bibnamefont{Verdozzi}}, \bibnamefont{and}
  \bibinfo{author}{\bibfnamefont{E.~K.~U.} \bibnamefont{Gross}},
  \bibinfo{journal}{Phys. Rev. Lett.} \textbf{\bibinfo{volume}{104}},
  \bibinfo{pages}{236801} (\bibinfo{year}{2010}).

\bibitem[{\citenamefont{Stefanucci and Kurth}(2011)}]{SteKur-PRL-11}
\bibinfo{author}{\bibfnamefont{G.}~\bibnamefont{Stefanucci}} \bibnamefont{and}
  \bibinfo{author}{\bibfnamefont{S.}~\bibnamefont{Kurth}},
  \bibinfo{journal}{Phys. Rev. Lett.} \textbf{\bibinfo{volume}{107}},
  \bibinfo{pages}{216401} (\bibinfo{year}{2011}).

\bibitem[{\citenamefont{Karlsson et~al.}(2011)\citenamefont{Karlsson,
  Privitera, and Verdozzi}}]{KarPriVer-PRL-11}
\bibinfo{author}{\bibfnamefont{D.}~\bibnamefont{Karlsson}},
  \bibinfo{author}{\bibfnamefont{A.}~\bibnamefont{Privitera}},
  \bibnamefont{and} \bibinfo{author}{\bibfnamefont{C.}~\bibnamefont{Verdozzi}},
  \bibinfo{journal}{Phys. Rev. Lett.} \textbf{\bibinfo{volume}{106}},
  \bibinfo{pages}{166401} (\bibinfo{year}{2011}).

\bibitem[{\citenamefont{Stoudenmire et~al.}(2012)\citenamefont{Stoudenmire,
  Wagner, White, and Burke}}]{StoWagWhiBur-PRL-12}
\bibinfo{author}{\bibfnamefont{E.}~\bibnamefont{Stoudenmire}},
  \bibinfo{author}{\bibfnamefont{L.~O.} \bibnamefont{Wagner}},
  \bibinfo{author}{\bibfnamefont{S.~R.} \bibnamefont{White}}, \bibnamefont{and}
  \bibinfo{author}{\bibfnamefont{K.}~\bibnamefont{Burke}},
  \bibinfo{journal}{Phys. Rev. Lett.} \textbf{\bibinfo{volume}{109}},
  \bibinfo{pages}{056402} (\bibinfo{year}{2012}).

\bibitem[{\citenamefont{Bergfield et~al.}(2012)\citenamefont{Bergfield, Liu,
  Burke, and Stafford}}]{BerLiuBurSta-PRL-12}
\bibinfo{author}{\bibfnamefont{J.~P.} \bibnamefont{Bergfield}},
  \bibinfo{author}{\bibfnamefont{Z.-F.} \bibnamefont{Liu}},
  \bibinfo{author}{\bibfnamefont{K.}~\bibnamefont{Burke}}, \bibnamefont{and}
  \bibinfo{author}{\bibfnamefont{C.~A.} \bibnamefont{Stafford}},
  \bibinfo{journal}{Phys. Rev. Lett.} \textbf{\bibinfo{volume}{108}},
  \bibinfo{pages}{066801} (\bibinfo{year}{2012}).

\bibitem[{\citenamefont{Ramsden and Godby}(2012)}]{RamGod-PRL-12}
\bibinfo{author}{\bibfnamefont{J.~D.} \bibnamefont{Ramsden}} \bibnamefont{and}
  \bibinfo{author}{\bibfnamefont{R.~W.} \bibnamefont{Godby}},
  \bibinfo{journal}{Phys. Rev. Lett.} \textbf{\bibinfo{volume}{109}},
  \bibinfo{pages}{036402} (\bibinfo{year}{2012}).

\bibitem[{\citenamefont{Buijse et~al.}(1989)\citenamefont{Buijse, Baerends, and
  Snijders}}]{BuiBaeSni-PRA-89}
\bibinfo{author}{\bibfnamefont{M.~A.} \bibnamefont{Buijse}},
  \bibinfo{author}{\bibfnamefont{E.~J.} \bibnamefont{Baerends}},
  \bibnamefont{and} \bibinfo{author}{\bibfnamefont{J.~G.}
  \bibnamefont{Snijders}}, \bibinfo{journal}{Phys. Rev. A}
  \textbf{\bibinfo{volume}{{40}}}, \bibinfo{pages}{4190}
  (\bibinfo{year}{1989}).

\bibitem[{\citenamefont{Filippi et~al.}(1994)\citenamefont{Filippi, Umrigar,
  and Taut}}]{FilUmrTau-JCP-94}
\bibinfo{author}{\bibfnamefont{C.}~\bibnamefont{Filippi}},
  \bibinfo{author}{\bibfnamefont{C.~J.} \bibnamefont{Umrigar}},
  \bibnamefont{and} \bibinfo{author}{\bibfnamefont{M.}~\bibnamefont{Taut}},
  \bibinfo{journal}{J. Chem. Phys.} \textbf{\bibinfo{volume}{100}},
  \bibinfo{pages}{1290} (\bibinfo{year}{1994}).

\bibitem[{\citenamefont{Gritsenko et~al.}(1995)\citenamefont{Gritsenko, van
  Leeuwen, and Baerends}}]{GrivanBae-PRA-95}
\bibinfo{author}{\bibfnamefont{O.~V.} \bibnamefont{Gritsenko}},
  \bibinfo{author}{\bibfnamefont{R.}~\bibnamefont{van Leeuwen}},
  \bibnamefont{and} \bibinfo{author}{\bibfnamefont{E.~J.}
  \bibnamefont{Baerends}}, \bibinfo{journal}{Phys. Rev. A}
  \textbf{\bibinfo{volume}{52}}, \bibinfo{pages}{1870} (\bibinfo{year}{1995}).

\bibitem[{\citenamefont{Gritsenko et~al.}(1996)\citenamefont{Gritsenko, van
  Leeuwen, and Baerends}}]{GrivanBae-JCP-96}
\bibinfo{author}{\bibfnamefont{O.~V.} \bibnamefont{Gritsenko}},
  \bibinfo{author}{\bibfnamefont{R.}~\bibnamefont{van Leeuwen}},
  \bibnamefont{and} \bibinfo{author}{\bibfnamefont{E.~J.}
  \bibnamefont{Baerends}}, \bibinfo{journal}{J. Chem. Phys.}
  \textbf{\bibinfo{volume}{104}}, \bibinfo{pages}{8535} (\bibinfo{year}{1996}).

\bibitem[{\citenamefont{Colonna and Savin}(1999)}]{ColSav-JCP-99}
\bibinfo{author}{\bibfnamefont{F.}~\bibnamefont{Colonna}} \bibnamefont{and}
  \bibinfo{author}{\bibfnamefont{A.}~\bibnamefont{Savin}}, \bibinfo{journal}{J.
  Chem. Phys.} \textbf{\bibinfo{volume}{{110}}}, \bibinfo{pages}{2828}
  (\bibinfo{year}{1999}).

\bibitem[{\citenamefont{Mori-Sanchez et~al.}(2009)\citenamefont{Mori-Sanchez,
  Cohen, and Yang}}]{MorCohYan-PRL-09}
\bibinfo{author}{\bibfnamefont{P.}~\bibnamefont{Mori-Sanchez}},
  \bibinfo{author}{\bibfnamefont{A.~J.} \bibnamefont{Cohen}}, \bibnamefont{and}
  \bibinfo{author}{\bibfnamefont{W.~T.} \bibnamefont{Yang}},
  \bibinfo{journal}{Phys. Rev. Lett.} \textbf{\bibinfo{volume}{{102}}},
  \bibinfo{pages}{066403} (\bibinfo{year}{2009}).

\bibitem[{\citenamefont{Wagner et~al.}(2012)\citenamefont{Wagner, Stoudenmire,
  Burke, and White}}]{WagStoBurWhi-PCCP-12}
\bibinfo{author}{\bibfnamefont{L.~O.} \bibnamefont{Wagner}},
  \bibinfo{author}{\bibfnamefont{E.~M.} \bibnamefont{Stoudenmire}},
  \bibinfo{author}{\bibfnamefont{K.}~\bibnamefont{Burke}}, \bibnamefont{and}
  \bibinfo{author}{\bibfnamefont{S.~R.} \bibnamefont{White}},
  \bibinfo{journal}{Phys. Chem. Chem. Phys.} \textbf{\bibinfo{volume}{{14}}},
  \bibinfo{pages}{8581} (\bibinfo{year}{2012}).

\bibitem[{\citenamefont{Levy}(1979)}]{Lev-PNAS-79}
\bibinfo{author}{\bibfnamefont{M.}~\bibnamefont{Levy}}, \bibinfo{journal}{Proc.
  Natl. Acad. Sci. U.S.A.} \textbf{\bibinfo{volume}{76}}, \bibinfo{pages}{6062}
  (\bibinfo{year}{1979}).

\bibitem[{\citenamefont{Seidl}(1999)}]{Sei-PRA-99}
\bibinfo{author}{\bibfnamefont{M.}~\bibnamefont{Seidl}},
  \bibinfo{journal}{Phys. Rev. A} \textbf{\bibinfo{volume}{{60}}},
  \bibinfo{pages}{4387} (\bibinfo{year}{1999}).

\bibitem[{\citenamefont{Seidl et~al.}(1999)\citenamefont{Seidl, Perdew, and
  Levy}}]{SeiPerLev-PRA-99}
\bibinfo{author}{\bibfnamefont{M.}~\bibnamefont{Seidl}},
  \bibinfo{author}{\bibfnamefont{J.~P.} \bibnamefont{Perdew}},
  \bibnamefont{and} \bibinfo{author}{\bibfnamefont{M.}~\bibnamefont{Levy}},
  \bibinfo{journal}{Phys. Rev. A} \textbf{\bibinfo{volume}{{59}}},
  \bibinfo{pages}{51} (\bibinfo{year}{1999}).

\bibitem[{\citenamefont{Seidl et~al.}(2000{\natexlab{a}})\citenamefont{Seidl,
  Perdew, and Kurth}}]{SeiPerKur-PRL-00}
\bibinfo{author}{\bibfnamefont{M.}~\bibnamefont{Seidl}},
  \bibinfo{author}{\bibfnamefont{J.~P.} \bibnamefont{Perdew}},
  \bibnamefont{and} \bibinfo{author}{\bibfnamefont{S.}~\bibnamefont{Kurth}},
  \bibinfo{journal}{Phys. Rev. Lett.} \textbf{\bibinfo{volume}{{84}}},
  \bibinfo{pages}{5070} (\bibinfo{year}{2000}{\natexlab{a}}).

\bibitem[{\citenamefont{Seidl et~al.}(2007)\citenamefont{Seidl, Gori-Giorgi,
  and Savin}}]{SeiGorSav-PRA-07}
\bibinfo{author}{\bibfnamefont{M.}~\bibnamefont{Seidl}},
  \bibinfo{author}{\bibfnamefont{P.}~\bibnamefont{Gori-Giorgi}},
  \bibnamefont{and} \bibinfo{author}{\bibfnamefont{A.}~\bibnamefont{Savin}},
  \bibinfo{journal}{Phys. Rev. A} \textbf{\bibinfo{volume}{{75}}},
  \bibinfo{pages}{042511} (\bibinfo{year}{2007}).

\bibitem[{\citenamefont{Gori-Giorgi and Seidl}(2010)}]{GorSei-PCCP-10}
\bibinfo{author}{\bibfnamefont{P.}~\bibnamefont{Gori-Giorgi}} \bibnamefont{and}
  \bibinfo{author}{\bibfnamefont{M.}~\bibnamefont{Seidl}},
  \bibinfo{journal}{Phys. Chem. Chem. Phys.} \textbf{\bibinfo{volume}{{12}}},
  \bibinfo{pages}{14405} (\bibinfo{year}{2010}).

\bibitem[{\citenamefont{R\"as\"anen et~al.}(2011)\citenamefont{R\"as\"anen,
  Seidl, and Gori-Giorgi}}]{RasSeiGor-PRB-11}
\bibinfo{author}{\bibfnamefont{E.}~\bibnamefont{R\"as\"anen}},
  \bibinfo{author}{\bibfnamefont{M.}~\bibnamefont{Seidl}}, \bibnamefont{and}
  \bibinfo{author}{\bibfnamefont{P.}~\bibnamefont{Gori-Giorgi}},
  \bibinfo{journal}{Phys. Rev. B} \textbf{\bibinfo{volume}{{83}}},
  \bibinfo{pages}{195111} (\bibinfo{year}{2011}).

\bibitem[{\citenamefont{Buttazzo et~al.}(2012)\citenamefont{Buttazzo, {De
  Pascale}, and Gori-Giorgi}}]{ButDepGor-PRA-12}
\bibinfo{author}{\bibfnamefont{G.}~\bibnamefont{Buttazzo}},
  \bibinfo{author}{\bibfnamefont{L.}~\bibnamefont{{De Pascale}}},
  \bibnamefont{and}
  \bibinfo{author}{\bibfnamefont{P.}~\bibnamefont{Gori-Giorgi}},
  \bibinfo{journal}{Phys. Rev. A} \textbf{\bibinfo{volume}{{85}}},
  \bibinfo{pages}{062502} (\bibinfo{year}{2012}).

\bibitem[{\citenamefont{Gori-Giorgi
  et~al.}(2009{\natexlab{a}})\citenamefont{Gori-Giorgi, Seidl, and
  Vignale}}]{GorSeiVig-PRL-09}
\bibinfo{author}{\bibfnamefont{P.}~\bibnamefont{Gori-Giorgi}},
  \bibinfo{author}{\bibfnamefont{M.}~\bibnamefont{Seidl}}, \bibnamefont{and}
  \bibinfo{author}{\bibfnamefont{G.}~\bibnamefont{Vignale}},
  \bibinfo{journal}{Phys. Rev. Lett.} \textbf{\bibinfo{volume}{{103}}},
  \bibinfo{pages}{166402} (\bibinfo{year}{2009}{\natexlab{a}}).

\bibitem[{\citenamefont{Liu and Burke}(2009)}]{LiuBur-JCP-09}
\bibinfo{author}{\bibfnamefont{Z.~F.} \bibnamefont{Liu}} \bibnamefont{and}
  \bibinfo{author}{\bibfnamefont{K.}~\bibnamefont{Burke}}, \bibinfo{journal}{J.
  Chem. Phys.} \textbf{\bibinfo{volume}{{131}}}, \bibinfo{pages}{124124}
  (\bibinfo{year}{2009}).

\bibitem[{\citenamefont{Levy and Perdew}(1985)}]{LevPer-PRA-85}
\bibinfo{author}{\bibfnamefont{M.}~\bibnamefont{Levy}} \bibnamefont{and}
  \bibinfo{author}{\bibfnamefont{J.~P.} \bibnamefont{Perdew}},
  \bibinfo{journal}{Phys. Rev. A} \textbf{\bibinfo{volume}{32}},
  \bibinfo{pages}{2010} (\bibinfo{year}{1985}).

\bibitem[{\citenamefont{Gori-Giorgi
  et~al.}(2009{\natexlab{b}})\citenamefont{Gori-Giorgi, Vignale, and
  Seidl}}]{GorVigSei-JCTC-09}
\bibinfo{author}{\bibfnamefont{P.}~\bibnamefont{Gori-Giorgi}},
  \bibinfo{author}{\bibfnamefont{G.}~\bibnamefont{Vignale}}, \bibnamefont{and}
  \bibinfo{author}{\bibfnamefont{M.}~\bibnamefont{Seidl}}, \bibinfo{journal}{J.
  Chem. Theory Comput.} \textbf{\bibinfo{volume}{{5}}}, \bibinfo{pages}{743}
  (\bibinfo{year}{2009}{\natexlab{b}}).

\bibitem[{\citenamefont{Mirtschink et~al.}()\citenamefont{Mirtschink, Seidl,
  and Gori-Giorgi}}]{MirSeiGor-XXX}
\bibinfo{author}{\bibfnamefont{A.}~\bibnamefont{Mirtschink}},
  \bibinfo{author}{\bibfnamefont{M.}~\bibnamefont{Seidl}}, \bibnamefont{and}
  \bibinfo{author}{\bibfnamefont{P.}~\bibnamefont{Gori-Giorgi}},
  \bibinfo{howpublished}{in preparation}.

\bibitem[{\citenamefont{Bednarek et~al.}(2003)\citenamefont{Bednarek, Szafran,
  Chwiej, and Adamowski}}]{BedSzaChwAda-PRB-03}
\bibinfo{author}{\bibfnamefont{S.}~\bibnamefont{Bednarek}},
  \bibinfo{author}{\bibfnamefont{B.}~\bibnamefont{Szafran}},
  \bibinfo{author}{\bibfnamefont{T.}~\bibnamefont{Chwiej}}, \bibnamefont{and}
  \bibinfo{author}{\bibfnamefont{J.}~\bibnamefont{Adamowski}},
  \bibinfo{journal}{Phys. Rev. B} \textbf{\bibinfo{volume}{{68}}},
  \bibinfo{pages}{045328} (\bibinfo{year}{2003}).

\bibitem[{\citenamefont{Casula et~al.}(2006)\citenamefont{Casula, Sorella, and
  Senatore}}]{CasSorSen-PRB-06}
\bibinfo{author}{\bibfnamefont{M.}~\bibnamefont{Casula}},
  \bibinfo{author}{\bibfnamefont{S.}~\bibnamefont{Sorella}}, \bibnamefont{and}
  \bibinfo{author}{\bibfnamefont{G.}~\bibnamefont{Senatore}},
  \bibinfo{journal}{Phys. Rev. B} \textbf{\bibinfo{volume}{{74}}},
  \bibinfo{pages}{245427} (\bibinfo{year}{2006}).

\bibitem[{\citenamefont{Almbladh and von Barth}(1985)}]{AlmBar-PRB-85}
\bibinfo{author}{\bibfnamefont{C.-O.} \bibnamefont{Almbladh}} \bibnamefont{and}
  \bibinfo{author}{\bibfnamefont{U.}~\bibnamefont{von Barth}},
  \bibinfo{journal}{Phys. Rev. B} \textbf{\bibinfo{volume}{31}},
  \bibinfo{pages}{3231} (\bibinfo{year}{1985}).

\bibitem[{\citenamefont{Levy et~al.}(1984)\citenamefont{Levy, Perdew, and
  Sahni}}]{LevPerSah-PRA-84}
\bibinfo{author}{\bibfnamefont{M.}~\bibnamefont{Levy}},
  \bibinfo{author}{\bibfnamefont{J.~P.} \bibnamefont{Perdew}},
  \bibnamefont{and} \bibinfo{author}{\bibfnamefont{V.}~\bibnamefont{Sahni}},
  \bibinfo{journal}{Phys. Rev. A} \textbf{\bibinfo{volume}{30}},
  \bibinfo{pages}{2745} (\bibinfo{year}{1984}).

\bibitem[{\citenamefont{Mendl and Lin}()}]{MenLin-arxiv-12}
\bibinfo{author}{\bibfnamefont{C.~B.} \bibnamefont{Mendl}} \bibnamefont{and}
  \bibinfo{author}{\bibfnamefont{L.}~\bibnamefont{Lin}},
  \bibinfo{howpublished}{arXiv:1210.7117}.

\bibitem[{\citenamefont{Seidl et~al.}(2000{\natexlab{b}})\citenamefont{Seidl,
  Perdew, and Kurth}}]{SeiPerKur-PRA-00}
\bibinfo{author}{\bibfnamefont{M.}~\bibnamefont{Seidl}},
  \bibinfo{author}{\bibfnamefont{J.~P.} \bibnamefont{Perdew}},
  \bibnamefont{and} \bibinfo{author}{\bibfnamefont{S.}~\bibnamefont{Kurth}},
  \bibinfo{journal}{Phys. Rev. A} \textbf{\bibinfo{volume}{{62}}},
  \bibinfo{pages}{012502} (\bibinfo{year}{2000}{\natexlab{b}}).

\end{thebibliography}

\end{document}